\documentclass{elsart}
\usepackage{epsfig}
\newcommand{\prl}[3]{Phys.\ Rev.\ Lett.\     {\bf  #1},  #2    (#3)}

\begin{document}
\begin{frontmatter}
%\preprint{BARI-TH/305-98}
\title{Wavelet Analysis of Blood Pressure Waves \\ 
	in Vasovagal Syncope}
\author{A. Marrone, A.D. Polosa, and G. Scioscia }
\address{Dipartimento di Fisica Universit\`a di Bari
 and Sezione INFN di Bari, \\
 Via Amendola 173, I-70126 Bari, Italy}
\author{S. Stramaglia}
\address{Istituto Elaborazione Segnali ed Immagini,\\
Consiglio Nazionale delle Ricerche,\\
Via Amendola 166/5, I-70126 Bari, Italy}
\author{A. Zenzola}
\address{Dipartimento di Scienze Neurologiche e 
	Psichiatriche Universit\`a di Bari, \\ 
	Piazza Giulio Cesare 11, I-70100 Bari, Italy}
\date{\rm BARI-TH/305-98}

%%%%%%%%%%%%%%%%%%%%%%% to be deleted
%\begin{center}
%\fbox{\large\bf DRAFT}
%\end{center}
%%%%%%%%%%%%%%%%%%%%%%%%%%%%%%%%%%%%%

\begin{abstract}
We describe the multiresolution wavelet analysis of blood pressure 
waves in vaso\-va\-gal-\-syncope affected patients compared 
with healthy people. We argue that there exist
discriminating criteria which allow us to isolate particular features, 
common to  syncope-affected patients sample,  indicating a tentative,
alternative diagnosis methodology for this syndrome.
We perform a throughout analysis both in the Haar 
basis and in a Gaussian one, with an attempt to grasp the
underlying dynamics.  
\end{abstract}
\end{frontmatter}

\newpage

%%%%%%%%%%%%%%%%%%%%%%%%%%%%%%%%%%%%%
\section{Introduction}
%%%%%%%%%%%%%%%%%%%%%%%%%%%%%%%%%%%%%
In recent years wavelet techniques have been successfully 
applied to a wide area of problems ranging from the image data 
analysis to the study of human biological rhythms~\cite{Th98}.
In the following we will make use of the so called
Discrete-Wavelet-Transform (DWT)   
to study  the temporal series generated from the human blood pressure
waves maxima, with particular attention  
to possible characteristic patterns
connected with Vasovagal Syncope (VS). The connection with
Fourier power spectra is put in evidence and the so called 
technique of
the Wavelet-Transform-Modulus-Maxima-Method (WTMM), 
is used~\cite{dna}.

VS is a sudden, rapid and reversing 
loss of consciousness, due to a reduction of cerebral blood 
flow~\cite{Ka95} attributable neither to cardiac structural 
or functional 
pathology, nor to neurological structural alterations, 
but due to a dysfunction of the cardiovascular control,
induced by that part of the   
Autonomic Nervous System (ANS) that regulates 
the arterial pressure~\cite{Ka95,Wo93}.
In normal conditions the arterial pressure is maintained 
at a constant level by  a negative feed-back 
mechanism localized in some nervous centers of the brain-stem. 
As a  consequence of a blood pressure variation, 
the ANS is able to restore the haemodynamic situation acting
on  heart and vases by means of two efferent pathways,
the vasovagal and sympathetic one, the former acting in the sense 
of a reduction of the arterial pressure, the latter in the opposite
sense~\cite{Gr66}.
VS consists of an abrupt fall of blood pressure 
corresponding to an acute haemodynamic
reaction produced by a sudden change in the activity of the ANS 
(an excessive enhancement of vasovagal outflow or a sudden decrease 
of sympathetic activity)~\cite{Ka95}.

VS is a quite common clinical problem and 
in the~$50\%$ of patients it is not diagnosed, being 
labelled as syncope of unknown origin, i.e. not necessarily 
connected to a dysfunction of the ANS\cite{Wo93,K95b,Ru95}. 
Anyway, a specific diagnosis of VS 
is practicable~\cite{K95b,Ka94}  with the help 
of the head-up tilt test (HUT)~\cite{Ke86}.
During this test the patient, positioned on a self-moving table, 
after an initial rest period in horizontal  position, 
is suddenly brought in vertical  position. 
Under these circumstances the ANS experiences a sudden stimulus 
of reduction of arterial pressure due to the shift of blood volume 
to inferior limbs. A wrong response to this stimulus 
can induce syncope behavior.

According to some authors, the positiveness of HUT means 
an individual predisposition toward VS\cite{Gr96}.
This statement does not find a general agreement because of 
the low reproducibility of the test~\cite{Ru96} in the same patient
and the extreme variability of the sensitivity in most 
of the clinical studies~\cite{Ka94}.
For this reason a long and careful clinical observation period is
needed to establish with a certain reliability whether
the patient is affected
by this syndrome.
In last years a large piece of work has been devoted to 
the investigation of signal patterns that could characterise
syncope-affected patients. This has been performed especially
by means of mathematical analyses of  arterial pressure and 
heart rate.
In particular the Fourier spectral analysis has shown 
to be unsuccessful for  this purpose~\cite{Wa61}.
In this paper we perform, by means of DWT, a new, detailed analysis 
of blood pressure waves of healthy people (controls) and 
syncope affected patients (positives) with the main intent 
to highlight all possible differences.
The positiveness of examined patients has been clinically established
after a long  observation period and also as a
consequence of repeated HUT tests.

The plan of the paper is as follows. In Section 2 we 
describe our data record, in Section 3 we  
give a short  mathematical introduction to Haar wavelet 
analysis with the aim of writing down the formulas that are
used in the subsequent Sections.  Section 4 is devoted 
to the results
of the DWT analysis while in Section 5 we investigate the possibility
of a scale-independent measure discriminating between healthy
and syncope affected patients.
Our conclusions are summarised in Section 6.
%%%%%%%%%%%%%%%%%%%%%%%%%%%%%%%%%%%%%%%
\section{Data Record}
%%%%%%%%%%%%%%%%%%%%%%%%%%%%%%%%%%%%%%%
%
%
\begin{figure}[t]
\begin{center}
   \epsfig{bbllx=0.5truecm,bblly=0.5truecm,bburx=17.5truecm,
   bbury=7.truecm,width=13.5truecm,clip=,file=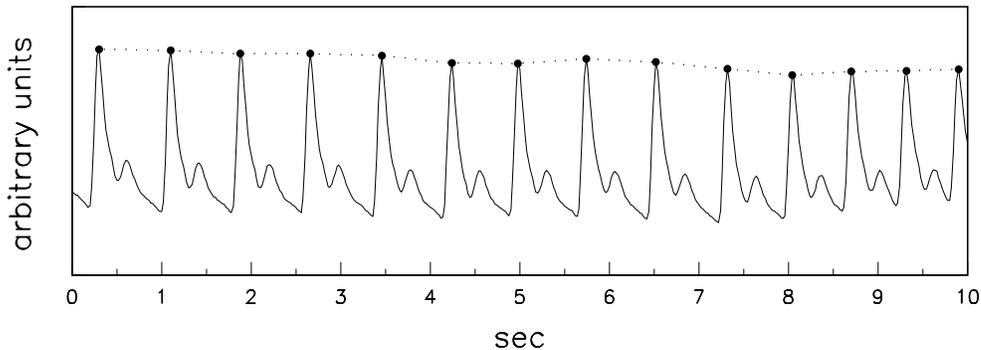}
	\caption{\small Blood pressure wave. The data we use
	in our analysis are the pressure heights of maxima.
	\label{fig:1}}
\end{center}
\end{figure}
Since the temporal behavior of blood pressure seems to be the most 
clinically relevant aspect to study VS, we extract  
the {\it heights} of blood pressure maxima from a recording 
period  twenty minutes long (which
is the better we can do for technical reasons) with the patient in 
horizontal rest  position. 
During this time the following biological signals of the subject
are recorded: E.C.G. (lead  D-II), E.E.G., the thoracic breath,
the arterial blood pressure (by means of a system 
{\em finapres~Ohmeda~2300}
Eglewood co. USA, measuring from the second finger of the left hand).
In Fig.~\ref{fig:1}  we show the pressure wave shape together
with the pressure maxima (circles) constituting the temporal series
we analyse.
Different individuals have different average values 
of pressure, therefore their blood wave signal has a different 
average height, but similar shape. 
What we have really care of, is the variation of height between 
neighbouring pressure maxima and not their absolute height values.
Anyway,  it can be realized that, in Haar basis, 
wavelet coefficients are rescaled derivatives 
of the function being analysed (see Sec. 3) so that 
undesired information is cancelled out.
The data set we analysed consists of 9 healthy people and 10 
syncope affected patients.
%%%%%%%%%%%%%%%%%%%%%%%%%%%%%%%%%%%%%
\section{The Haar wavelet analysis}
%%%%%%%%%%%%%%%%%%%%%%%%%%%%%%%%%%%%%
In the present Section we give a brief account of wavelet 
mathematical aspects that are relevant to our objectives~\cite{Rass}.
The most striking difference between Fourier and DWT
decomposition is that the last allows for a projection 
on modes simultaneously localized in 
both time and frequency space, up to the limit of classical 
uncertainty relations. 
Unlike the Fourier bases, which are delocalised for definition, 
the DWT bases have compact spatial support, therefore being 
particularly suitable for the study of signals which are known 
only inside a limited temporal window. 

The Haar wavelet is historically the first basis  
  introduced for wavelet analysis
and for many practical purposes it is the simplest 
to be used in applications. 
Let us consider a function $f(x)$, defined in $[0,L]$, 
representing some data. 
This function is typically known with some finite
resolution $\Delta x$ and it can be represented as 
an histogram having $2^{m}$ bins in such a way that:
\begin{equation}
\frac{L}{2^{m}} \geq \Delta x .
\end{equation}
Each bin is labelled by an integer $n$ running from $0$ to $2^m-1$. 
We can now define:
\begin{equation}
f_{m,n}(x)=f(x),\;\;\; x\in \left[ \frac{L}{2^{m}} n,
	\frac{L}{2^{m}}(n+1)\right].
\end{equation}
Obviously the following relation holds:
\begin{equation}
f(x)=\sum_{n=0}^{2^{m}-1} f_{m,n}(x) \phi \left( 
\frac{2^{m}}{L}x-n \right)=f^{(m)}(x) ,
\end{equation}
where  $\phi$ is defined  as :
\begin{equation}
\phi(s)  =  \left\{
 \begin{array}{lcl}
  1 & \;\;\;\; & {\rm if} \;\, 0  \leq s \leq 1 \\
  0 &          & {\rm elsewhere}\\
 \end{array}
 \right. \,  .	
\end{equation}
By   $f^{(m)}(x)$ we mean ``$f$  at scale $m$''.
Let us consider now a roughening of $f^{(m)}$:
\begin{equation}
f^{(m-1)}(x)= \sum_{n=0}^{2^{m-1}-1} f_{m-1,n}(x) 
	\phi \left(\frac{2^{m-1}}{L}x-n\right) ,
\end{equation}
where, as it's  easy to check, we have:
\begin{equation}
f_{m-1,n}(x) =\frac{1}{2}(f_{m,2n}(x)  + f_{m,2n+1}(x)) .
\end{equation}
$f^{(m-1)}(x)$ contains less information 
than $f^{(m)}(x)$, so, in order to recover the information 
that has been lost, we should be able to calculate the difference 
function $f^{(m)}(x)-f^{(m-1)}(x)$. 
Let us call this difference function $W^{(m-1)}(x)$. 
It can be shown (see~\cite{Rass}) that:
\begin{equation}
W^{(m-1)}(x)=\sum_{n=0}^{2^{m-1}-1} W_{m-1,n}(x) \psi 
	\left(\frac{2^{m-1}}{L}x-n \right) ,
\end{equation}
where:
\begin{equation}
W_{m-1,n}(x)=\frac{1}{2}(W_{m,2n}(x)-W_{m,2n+1}(x)) ,
\end{equation}
and:
\begin{equation}
\psi(s)  =  \left\{
 \begin{array}{lcl}
  1 & \;\;\;\; & {\rm if} \;\, 0  \leq s < 0.5 \\
 -1 &          & {\rm if} \;\, 0.5\leq s \leq 1\\
 \end{array}
 \right. 
\end{equation}
being $\psi(s)$ zero outside the indicated ranges.
$\phi$ and $\psi$ are respectively known as mother and father 
functions and the coefficients $f_{m,n}$, $W_{m,n}$  
are the mother and father (or wavelet) coefficients. 
Mother and father functions, taken  together, generate a compactly
supported orthogonal basis . 
According to our definition of difference function, 
it is straightforward to observe that:
\begin{equation}
f^{(m)}(x)=f^{(0)}(x)+W^{(0)}(x)+......+W^{(m-1)}(x).
\end{equation}
Analysing a signal which oscillates around an average value 
we realize that, for all practical purposes, $f^{(0)}(x)=0$, 
so that:
\begin{equation} 
f(x)=\sum_{j=0}^{m-1}\sum_{n=0}^{2^{j}-1} W_{j,n}(x) \psi 
	\left(\frac{2^{j}}{L}x-n \right) ,
\end{equation}
where summing on all $m$'s corresponds to looking at function $f$ 
at all possible scales.
This is the wavelet representation of $f(x)$ provided 
by the Haar basis. 
What we learn is that the difference functions $W$ enable us 
to project the function $f(x)$ on a new basis set. 
Furthermore, the orthogonality properties of mother and father 
function allow  to write down the coefficients of the
Discrete Wavelet Transform. 
A particular choice of normalization gives:
\begin{equation}
W_{m,n} = 2^{-m^{\prime}/2}\,\sum_{i=0}^{L-1} f_i 
\psi(2^{-m^{\prime}} i - n) , 			\label{eqn:coe}
\end{equation}
where $m^{\prime}=10-m$ and, for our scopes, $L=2^{10}$ 
i.e. the total number 
of pressure wave maxima $f_i$ in our data record,
constituting the function that we want to study.  
Therefore, substituting $m^{\prime} \rightarrow m$, 
we have:
\begin{equation}
W_{10-m,n} = 2^{-m /2}\,\sum_{i=0}^{L-1} f_i \psi(2^{-m } i - n) .
\end{equation}
Let us call $W_{10-m,n}=W^{\prime}_{m,n}$ in the following.
The variability of the  wavelet coefficients for each pressure wave
has been parameterized at the different scales 
(different values of $m$) by means of their standard deviations:
\begin{equation}
\sigma(m) = \left[ \frac{1}{N-1} \sum_{n=0}^{N-1} 
	(W^{\prime}_{m,n} - \langle W^{\prime}_{m,n} \rangle 
	)^2 \right]^{\frac{1}{2}} ,   \label{eqn:sig}
\end{equation}
where $N$ is the number of wavelet coefficients 
at a given scale $m$ ($N=L/2^m$).
%%%%%%%%%%%%%%%%%%%%%%%%%%%%%%%%%%%%%
\section{Results}
\begin{figure}[t]
\begin{center}
   \epsfig{bbllx=0.truecm,bblly=0.5truecm,bburx=18.truecm,%
	bbury=21.truecm,height=12.truecm,clip=,file=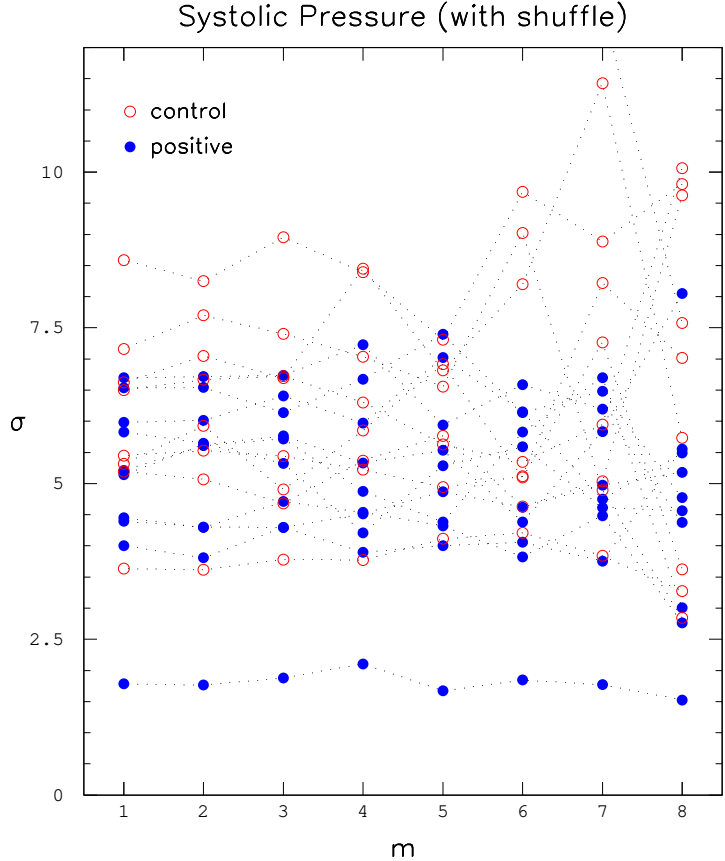}
	\caption{\small Standard deviations of the wavelet 
		coefficients of the Systolic  Pressure 
		computed after a randomly reordering
		of the pressure wave maxima. \label{fig:3}}
\end{center}
\end{figure}

%%%%%%%%%%%%%%%%%%%%%%%%%%%%%%%%%%%%%
Basically, all we need for our purposes, is the expression of the
fluctuation in Eq.~(\ref{eqn:sig}).
Our main results are   obtained
examining  plots of scale $m$ versus $\sigma(m)$ in which 
the function $f_i$ represents the height of maxima 
of systolic/diastolic blood pressure waves
of healthy and syncope-affected patients recorded. 
The variable $i$ is a time variable. We assume that 
the time interval separating two consecutive pressure maxima is, 
in standard conditions, almost equal to a certain
average value that we could establish case by case.
This does not spoil our analysis of the temporal information 
which actually plays a central role. 
Our results are in fact definitely dependent on the order
with which blood pressure maxima, extracted from data, 
are disposed in sequence and then analysed. 
This emerges clearly from the following considerations. 
\begin{figure}[t]
\begin{center}
   \epsfig{bbllx=0.truecm,bblly=0.8truecm,bburx=18.truecm,%
	bbury=18.truecm,height=12.truecm,clip=,file=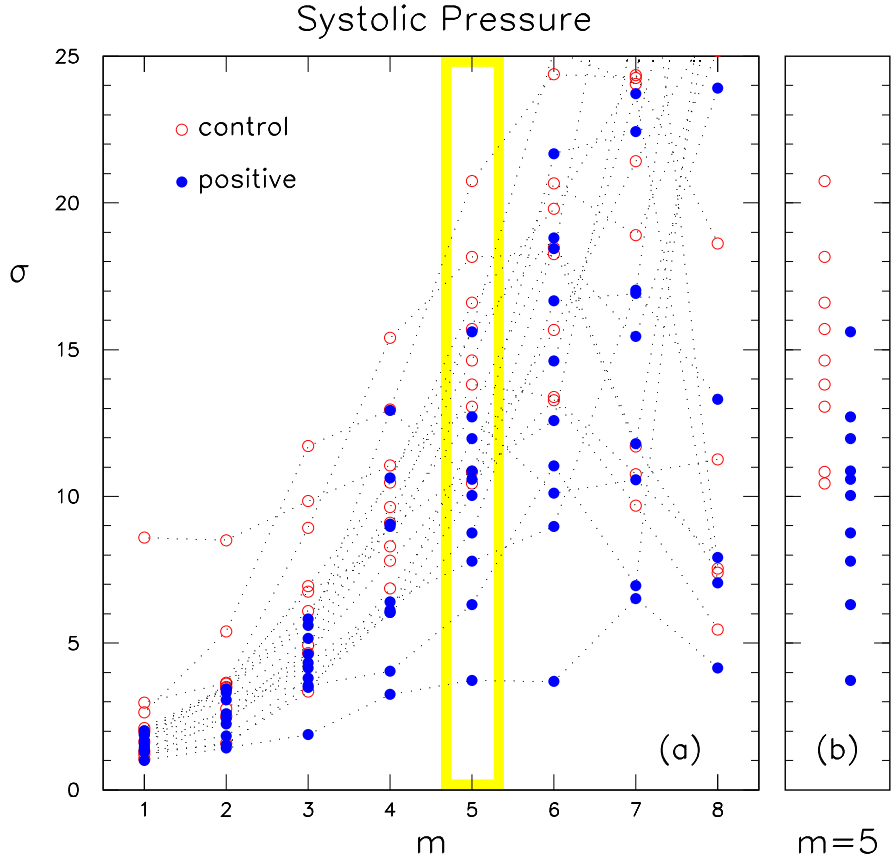}
	\caption{(a) Standard deviations of the wavelet coefficients 
		[Eq.~(\ref{eqn:sig})]
		of the Systolic  Pressure in syncope-affected 
		patients (positives) and healthy people (controls).
		(b) Quite evident separation among positives
		and controls at $m=5$. \label{fig:2}}
\end{center}
\end{figure}
Figure~\ref{fig:2}(a) shows the scale dependence of the fluctuations 
(see Eq.~(14)) computed for the systolic pressures of our data set.
A common trend of $\sigma$'s seems to emerge: the fluctuations
of healthy people (open circles) are, at each scale, slightly
higher than those of patients with syncope clinical behavior.
In particular, a quite clear distinction appears traced 
in correspondence of the $m=5$ scale value, as can be seen 
in Fig.~\ref{fig:2}(b).
For smaller values of $m$ there is no sensitive distinction 
between controls and positives, while, for bigger values 
our analysis begins to be less and less significant 
due to the narrowness of our temporal observation window.
This distinction is instead lost 
when a randomly chosen temporal rearrangement of pressure maxima 
is analysed, as it is shown in Fig.~\ref{fig:3}.  

If we interpret this analysis as a possible discriminating test, 
and if we trace a virtual threshold line in correspondence of 
$\sigma=12.9$, we 
observe that the test would have sensitivity of $90\%$ and 
a specificity of $78\%$.
To give a much more quantitative meaning to results 
of Fig.~\ref{fig:2}(b)
we perform a statistical test: the Wilcoxon-Mann-Whitney (WMW) 
test~\cite{libro}. 
This  is aimed to check the hypothesis 
that our two samples, controls and positives, have been drawn from 
populations  with the same continuous distribution function.
The WMW test gives to this hypothesis the $3.5\times10^{-3}$ 
probability, i.e.
the statistical hypothesis 
is rejectable at the level of significance of $1\%$.

These empirical observations have been subject of a further study
focused to understanding why $m=5$ is the relevant discriminating 
scale or possibly the onset of a discriminating region 
between controls and positives.
The result of this investigation is that $m=5$ separation 
corresponds to the fact that positives must have in their Fourier 
power spectrum, relative to the above-mentioned pressure signal, 
an ``hidden'' suppression of a certain low frequency range
(roughly centered around $0.02\div 0.03$~Hz, being $m=5$ 
the resolution window corresponding to about $2^5$ pressure maxima).
A simple, qualitative, explanation of these results 
is the following.\newline 
If one calculates, with the help of a computer,
Haar wavelet coefficients $W_{m,n}$ of a sinus (or cosinus) 
function using Eq.~(\ref{eqn:coe}), he will soon realize that 
what is obtained are derivatives of sinus (or cosinus) rescaled 
on the $x$-axis. Passing from an $m$ value to the next $m+1$ value,
a derivative is made and $x$ units are scaled 
by a dilation factor $a<1$. 
This means that, passing from an $m$ to the next, 
the sinusoidal function will have a greater fluctuation 
over the zero average value (the area below the curve must remain 
the same after the rescaling of coordinates).
According to what has just been said, a low frequency sinusoidal 
function, with a small weight coefficient,
will start fluctuating in a sensitive way 
only going up with $m$ values.
\begin{figure}[t]
\begin{center}
   \epsfig{bbllx=0.0truecm,bblly=1.3truecm,bburx=12.truecm,
	bbury=9.2truecm,height=9.truecm,clip=,file=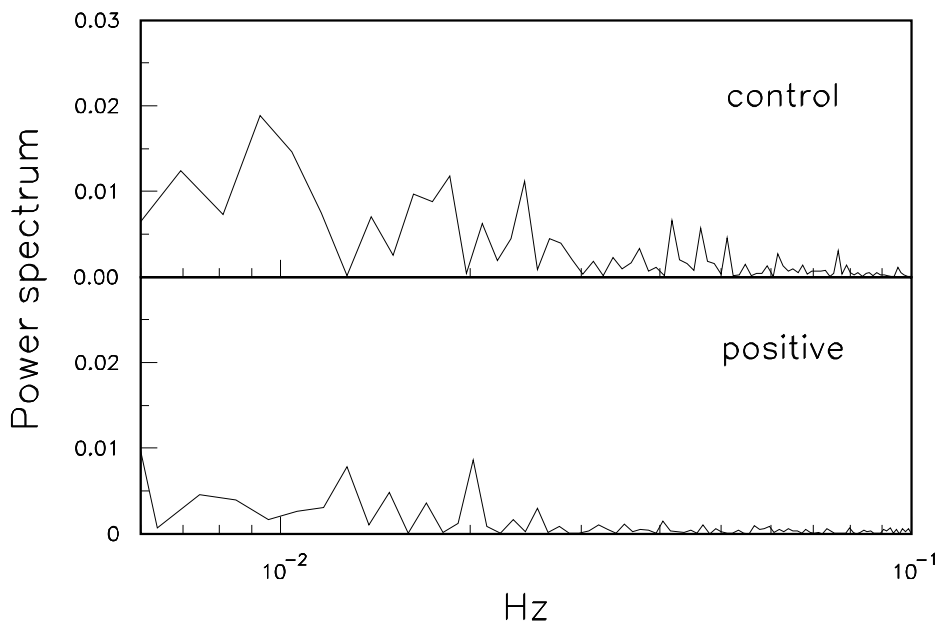}
	\caption{\small Low frequency power spectrum for a
	control and a positive. A similar behavior holds
	for all our data records.\label{fig:4}}
\end{center}
\end{figure}
As known from Fourier analysis, each function, under very general
conditions, can be written down as a sum 
of weighted sinusoidal functions.
If we take two functions differing in their Fourier coefficient 
values only in a common  low frequency range and 
we calculate their $\sigma (m)$ for different $m$'s, 
we expect that the function having the smaller weight coefficients 
in the low frequency domain will also have the smaller 
$\sigma$ values, for certain particular $m$ values, 
with respect to the other. 
For greater $m$'s, the depressed low frequency harmonics 
begin to fluctuate in a more sensitive way, so that we can 
foresee again a decreasing in the differences between 
$\sigma(m)$ values belonging to the two mentioned functions.

In Fig.~\ref{fig:4} we  show the Fourier power spectrum in the 
frequency domain of interest. We find  the predicted
low frequency depression
in positives.
The Fourier analysed series is again the systolic 
blood pressure maxima succession. 
The peaks around $0.05$~Hz which are visible for 
the displayed control power spectrum  are 
not systematically present in all our controls data set, 
while the above quoted difference between controls and positives 
in the $0.02$~Hz range is systematically present in our data records.  
A Fourier analysis of the entire pressure wave gives a very 
complicated  power spectrum and consequently it is almost 
impossible to make significant comparisons between 
positives and controls. 
Even if this result seems to be encouraging, it has a main drawback.
The $m=5$ channel may  strictly be connected to the wavelet basis 
set used and to the particular group of subjects~\cite{hestanley}. 
This at least requires  further 
investigations possibly aimed to  highlight features being somehow  
``universal", i.e. independent on individuals, but due to the
underlying dynamics of the ANS regulation. 
This is the scope of the next Section.
\begin{figure}[t]
\begin{center}
   \epsfig{bbllx=0.truecm,bblly=0.8truecm,bburx=18.truecm,%
	bbury=18.truecm,height=12.truecm,clip=,file=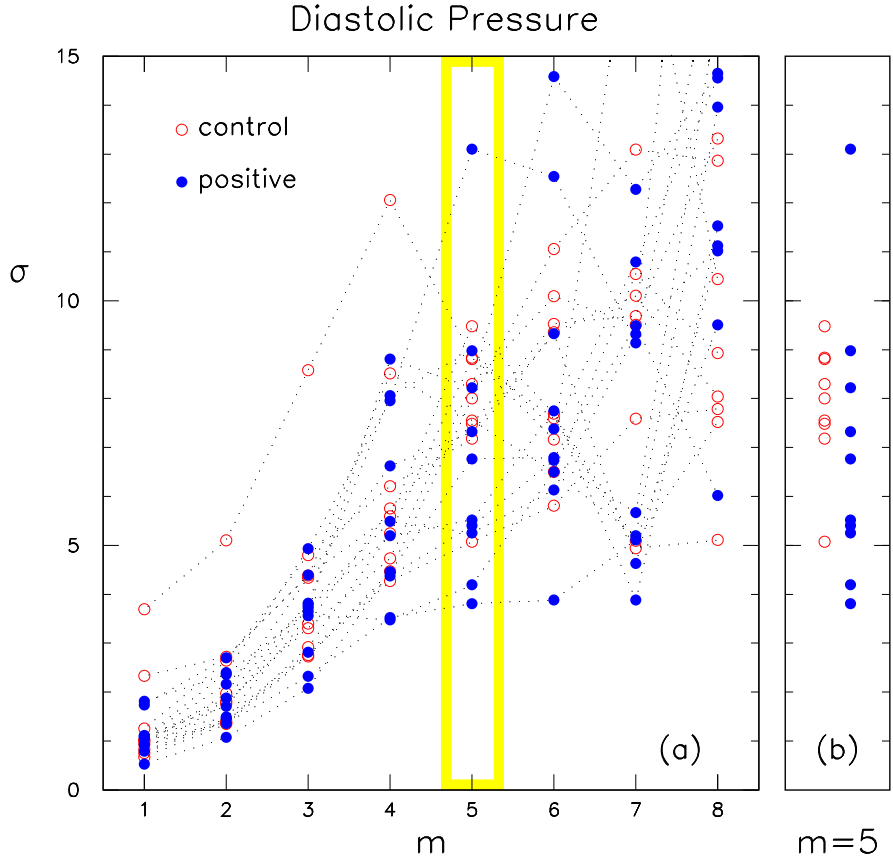}
	\caption{\small As in Fig.~\ref{fig:2} but for the 
		diastolic  pressure. \label{fig:5}}
\end{center}
\end{figure}

An analysis similar to that carried out for systolic pressure waves
has been performed for diastolic pressure waves 
belonging to the same patients and healthy subjects. 
As it is shown in Fig.~\ref{fig:5}(a), a clear evidence 
for separation between controls and positives is lost 
even if, at $m=5$~(Fig.~\ref{fig:5}(b)), controls tend
to accumulate towards higher $\sigma$-values than positives.
The WMW test, applied to $\sigma$'s in Fig.~\ref{fig:5}(b),
gives now a probability of  $8.2\%$.  
This loss of sensitivity may be due to the shorter  
variability range of the wavelet coefficients 
of the diastolic pressures with respect to the systolic ones, 
as can be extrapolated comparing the vertical scales
of Fig.~\ref{fig:5} with respect to those in Fig.~\ref{fig:2}.
A physiological grounded understanding of this phenomenon is at 
the moment lacking.

%%%%%%%%%%%%%%%%%%%%%%%%%%%%%%%%%%%%%
\section{Scale independent measures}
\begin{figure}[t]
\begin{center}
   \epsfig{bbllx=0.truecm,bblly=0.5truecm,bburx=12.truecm,
   bbury=4.5truecm,height=4.5truecm,clip=,file=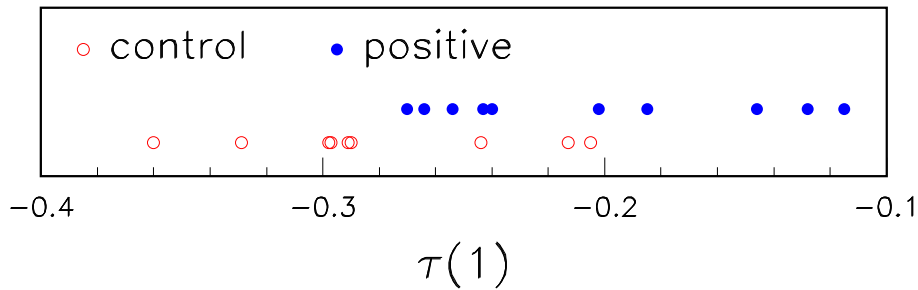}
	\caption{\small $\tau(q=1)$ exponents, extracted from our
	systolic  pressure record, are shown.\label{fig:6}}
\end{center}
\end{figure}

%%%%%%%%%%%%%%%%%%%%%%%%%%%%%%%%%%%%%
With an approach similar to that proposed by~\cite{hestanley} we 
investigate on the possibility to set us free from the  $m=5$
discriminating scale in order to find some ``universal" features
labelling our two distinct classes of subjects.
We have calculated the sum $Z_q$ of the $q$th moments of 
the coefficients of the wavelet transform, defined as~\cite{dna}:
\begin{equation}
Z_q(s)=\sum_{n=1}^{L}({\rm max} |W_{s,n}|)^q,
\end{equation}
where $s$ is a continuously varying scale.
Wavelet coefficients $W_{s,n}$ are  calculated using a formula 
slightly different from that in Eq. (12):
\begin{equation}
 W_{s,n}=\frac{1}{s}\sum_{i=1}^{L} f_i \psi\left(\frac{i-n}{s}\right),
\end{equation}
where now $n$ ranges in the interval $[1,2,3,...,L]$ 
and the $\psi$ function is
the third derivative of the Gaussian 
$\psi(x)=\frac{d^3}{dx^3} e^{-x^2/2}$.
For a fractal signal $f_i$, the following scaling law is expected:
\begin{equation}
Z_q(s)\sim s^{\tau(q)}.
\end{equation}
A measure of $\tau(q)$, obtained through log-log plots of
$Z_q(m)$~vs.~$m$ for each $q$, could be
interpreted as a scale independent measure characterising the
unknown underlying dynamics of the ANS regulation of blood pressure.
Our goal has been to find out a certain value of $q$ at which 
appears a significant difference between controls and positives. 
Using our data record with systolic pressure maxima, 
we find that the $\tau(q=1)$ acts as a discriminating
parameter (see Fig.~\ref{fig:6}) while with other $q$ values we 
do not succeed in obtaining  equally convincing results. 
Again we perform a WMW test obtaining a probability 
of $4.5\times 10^{-3}$
that the two sets of points displayed in Fig.~\ref{fig:6} belong 
to populations with the same continuous distribution function.
The same test repeated for the case of diastolic  
pressures (not shown), gives a
probability of $8.3\%$, indicating again 
a less reliability of this data set for our scopes.
Varying the wavelet basis used, we get
very different $Z_q(s)$ functions. 
The quoted scaling behavior in Eq. (17)
seems possible only if we use, as wavelet basis, the third derivative
of the Gaussian. 
With other bases, such as the Haar basis or the Daubechies one, this
scaling  is absent.
On the other hand a $\sigma(s)$~vs.~$s$ plot drawn with 
$\sigma(s)$ values calculated  in Gaussian basis, 
is completely ineffective to find some separation trend 
at any $s$ scale.  
%

%%%%%%%%%%%%%%%%%%%%%%%%%%%%%%%%%%%%% 
\section{Conclusions and perspectives}
%%%%%%%%%%%%%%%%%%%%%%%%%%%%%%%%%%%%%
We are aware that the analyses here suggested are far to 
be used as an operative diagnosing method of VS 
but we show that they strongly suggest 
some directions to look at. 
We are not sure that a scale independent measure, such 
that illustrated in Section 5, can give deeper insights about
the problems of VS diagnosis and of understanding  the dynamical
behavior of ANS than $\sigma(m)$~vs.~$m$ plots, illustrated 
in Section 4. This arises from the fact that the measure
described  is still linked to the choice of a particular wavelet
basis, as is $m$ in the analysis of Section 4.
According to us there is a sort of complementarity between the two 
described approaches which also emerges in the 
impressive correspondence between results of WMW tests
when applied to $\sigma$'s in Haar basis and $\tau$'s 
in Gaussian basis.   
\par\noindent
\ack{The authors are grateful to Doc. M.~Osei~Bonsu for giving us 
the possibility to access at the not elaborated 
blood pressure data.
A.~D.~P. acknowledges Prof. R.~Gatto for his long and 
kind hospitality at the University of Geneva and Commission 4 of INFN
for supporting in part this visiting period.}


\newpage

\begin{thebibliography}{99}

\bibitem{Th98}	S.~Thurner, M.C.~Feuerstein, and M.C.~Teich, 
		\prl{80}{1544}{1998}; Y.~Ashkenazy {\em et al.}, 
		{\em Fractals}~6(3), 197 (1998).

\bibitem{dna} 	A.~Arneodo, Y.~d'Aubenton-Carafa, E.~Barcy, 
		P.V.~Graves, J.F.~Muzy, C.~Thermes, 
		Physica D {\bf 96}, 291 (1996).
	      
\bibitem{Ka95}	H.~Kauffman, {\em Neurology}~45~(suppl. 5), 12~(1995).

\bibitem{Wo93}	D.A.~Wolfe {\em et al.}, 
		{\em Am. Fam. Physician}~47(1), 149~(1993).

\bibitem{Gr66}	R.~Greger, U.~Windhorst, {\em Comprehensive human
		physiology}, ed. Springer-Verlag Berlin 
		Heidelberg 1966, vol.~2, pag.~1995.

\bibitem{K95b}	W.N.~Kapoor, {\em Cliv. Clin. J. Med.}~62(5),
		305~(1995).

\bibitem{Ru95}	G.A.~Ruiz {\em et al.}, {\em Am. Heart J.}~130, 
		345~(1995).

\bibitem{Ka94}	W.N.~Kapoor, {\em Am. J. Med.}~97,~78~(1994).

\bibitem{Ke86}	R.A.~Kenny {\em et al.}, {\em Lancet}~14, 1352~(1986).

\bibitem{Gr96}	B.P.~Grubb, D.~Kosinski, {\em Current Opinion
		Cardiology}~11, 32~(1996); 
		R.~Sheldon {\em et al.}, {\em Circulation}~93,
		973~(1996).


\bibitem{Ru96}	G.A.~Ruiz {\em et al.}, 
		{\em Clin. Cardiol.}~19, 215~(1996).


\bibitem{Wa61}	A.~Malliani {\em et al.}, {\em Circulation}~84, 
		482~(1991).

\bibitem{Rass} 	R.A.~Gopinath {\em et al.}, 
		{\em Introduction to Wavelets and Wavelet 
		Transforms: a Primer}, Prentice Hall 1997;
               	G.~Kaiser, {\em A Friendly Guide to Wavelets}, 
		Birkhauser 1994;
               	L.Z.~Fang, J.~Pando, Report astro-ph/9701228,
		to appear in the Proceedings of the 5th Erice
		Chalonge School on Astrofundamental Physics,
		N.~S\'anchez and A.~Zichichi eds., 
		World Scientific, 1997.

\bibitem{libro} M.~Fisz, {\em Probability Theory and 
		Mathematical Statistics},
                Krieger Publishing Company, 3rd ed., 1980.      

\bibitem{hestanley} L.S.~Nunes Amaral, A.L.~Goldberger, 
		P.Ch.~Ivanov and H.E.~Stanley, \prl{81}{2388}{1998}.

\end{thebibliography}
\end{document}